\documentclass{article} 

\usepackage{iclr2022_conference,times}
\usepackage{color}

\usepackage{xspace} 
\newcommand{\eg}{\hbox{{e.g.}}\xspace}
\newcommand{\ie}{\hbox{{i.e.}}\xspace}
\usepackage{array,multirow,makecell}
\usepackage{mathrsfs}
\usepackage{float}
\usepackage{graphicx}
\usepackage{amsmath}
\usepackage{array}
\usepackage[english]{babel}
\usepackage{mdwmath}
\usepackage[ruled, linesnumbered]{algorithm2e}
\usepackage{adjustbox}
\usepackage{bbm}
\usepackage{mdwtab}
\usepackage{tabularx}
\newcolumntype{P}[1]{>{\centering\arraybackslash}p{#1}}
\newcolumntype{M}[1]{>{\centering\arraybackslash}m{#1}}
\usepackage{amssymb}
\usepackage{textcomp}
\usepackage{booktabs}
\usepackage{wrapfig}
\usepackage{multirow}
\usepackage[normalem]{ulem}
\useunder{\uline}{\ul}{}

\usepackage{amsmath,amsfonts,bm}

\def\eqref#1{equation~\ref{#1}}

\def\1{\bm{1}}

\DeclareMathAlphabet{\mathsfit}{\encodingdefault}{\sfdefault}{m}{sl}
\SetMathAlphabet{\mathsfit}{bold}{\encodingdefault}{\sfdefault}{bx}{n}

\usepackage{hyperref}
\usepackage{url}

\title{Fingerprinting Multi-exit Deep Neural Network Models via Inference Time}

\author{Tian Dong$^1$,~Han Qiu$^2$\thanks{qiuhan@tsinghua.edu.cn}~,~Tianwei Zhang$^3$,~Jiwei Li$^4$,~Hewu Li$^2$,~and~Jialiang Lu$^1$
\\$^1$Shanghai Jiao Tong University, China 
\\$^2$Tsinghua Univeristy, China 
\\$^3$Nanyang Technological University, Singapore
\\$^4$Shannon.AI, Zhejiang Univeristy, China
}
\vspace{-3em}

\newcommand{\norm}[1]{\left\lVert#1\right\rVert}
\begin{document}

\maketitle

\begin{abstract}

Transforming large deep neural network (DNN) models into the multi-exit architectures can overcome the overthinking issue and distribute a large DNN model on resource-constrained scenarios (e.g. IoT frontend devices and backend servers) for inference and transmission efficiency. Nevertheless, intellectual property (IP) protection for the multi-exit models in the wild is still an unsolved challenge. Previous efforts to verify DNN model ownership mainly rely on querying the model with specific samples and checking the responses, e.g., DNN watermarking and fingerprinting. However, they are vulnerable to adversarial settings such as adversarial training and are not suitable for the IP verification for multi-exit DNN models. In this paper, we propose a novel approach to fingerprint multi-exit models via \emph{inference time} rather than \emph{inference predictions}. Specifically, we design an effective method to generate a set of fingerprint samples to craft the inference process with a unique and robust inference time cost as the evidence for model ownership. We conduct extensive experiments to prove the uniqueness and robustness of our method on three structures (ResNet-56, VGG-16, and MobileNet) and three datasets (CIFAR-10, CIFAR-100, and Tiny-ImageNet) under comprehensive adversarial settings.

\end{abstract}

\vspace{-1em}
\section{Introduction}
\vspace{-1em}

Nowadays, deep neural networks (DNNs) tend to become more and more complex with the rapidly increasing number of layers and parameters~\citep{GPT3_cost} to achieve better performance. 
However, huge DNN models introduce issues like overthinking~\citep{huang2017multi} and are hard to deploy in resource-constrained scenarios like edge computing devices. 
One promising approach is to transform a large trained DNN model into a \textit{multi-exit} model. 
A multi-exit model is composed of a backbone model (\ie the large trained DNN model) and multiple internal classifiers (ICs) located at the end of some hidden layers of the backbone model~\citep{huang2017multi,kaya2019shallow}. 
The backbone model is used for feature extraction and the ICs allow samples to be predicted and to exit at an early layer of the model based on a tunable early-exit criteria. 
This multi-exit model structure can effectively avoid overthinking issues. 
Moreover, by fragmenting the multi-exit models according to their ICs, a few layers can be deployed on edge devices for some samples' inference and early exit to save the energy and transmission cost~\citep{hong2020panda}. 

However, how to protect the intellectual property (IP)~\citep{tramer2016stealing, orekondy2019prediction} of these multi-exit DNN models remains an unsolved challenge. 
In this paper, we name the initial DNN model as the \textbf{target model} and the model waiting for ownership verification as the \textbf{suspicious model}. 
Traditional IP protection methods on DNN models mainly rely on the remote query for inference with some special samples and verify the ownership by their predictions on suspicious models. 
For instance, DNN watermarking techniques~\citep{zhang2018protecting} aim to inject backdoors~\citep{li2020backdoor} into DNN models by modifying the training datasets or model parameters such that the backdoor samples can be used to verify the ownership by inference results. 

Later on, a more advanced DNN fingerprinting concept is proposed based on verifying model IP via its natural feature without modifying the model~\citep{cao2021ipguard}. 
Compared with DNN watermarking, DNN fingerprinting has various advantages such as no loss on the accuracy, availability for off-the-shelf models, etc. 
However, although with these good visions, existing DNN fingerprinting methods~\citep{conferrable, characteristic_example} still follow a similar way of generating adversarial examples (AE) by just modifying either perturbation bounds~\citep{cao2021ipguard} or transferability~\citep{conferrable}. 
Although they can verify IP against model compression, pruning, and fine-tuning, it is pointed out that they are vulnerable to adversarial training~\citep{conferrable}.

These DNN fingerprinting techniques cannot be used for multi-exit models due to the following two vulnerabilities. 
(1) Attackers may modify the multi-exit model structures by adding or removing ICs, retraining ICs, or changing the early-exit criteria to manipulate the prediction accuracy to compromise the fingerprinting prediction. 
Note that the training cost for the ICs is significantly less than training the backbone DNN model, hence, it is feasible for attackers to modify or re-train the ICs to avoid the IP verification. 
(2) The backbone DNN model for some multi-exit structures such as SDN~\citep{kaya2019shallow} is known to attackers, so they can directly modify the backbone DNN models with adversarial training to mitigate the AE-based fingerprinting techniques. 

In this work, we aim to first answer the following question: \textit{how to design an effective fingerprinting technique for the multi-exit models?} 
Here we define the \textit{effective} with two specific requirements: \textit{uniqueness} refers to low chances of mis-verifying similar but \textbf{independent models} (\eg models trained with the same dataset and structure but different initial seeds); \textit{robustness} aims to overcome the above vulnerabilities of existing DNN fingerprinting methods. 
The key insight of our idea is to \textit{fingerprint the multi-exit models via inference time of fingerprint samples}. 
Specifically, our idea is to craft a set of fingerprint samples that will exit at certain layers of the target model but behave normally on independent models considering the inference time to guarantee the \textit{uniqueness}. 
Then, since our fingerprinting does not rely on the prediction results, it is not affected by modifications on the multi-exit models such as modifying ICs to manipulate prediction accuracy or modifying the backbone models. 
Therefore, our fingerprinting method can realize \textit{robustness}.

In summary, our contribution can be summarized as follows. 
(1) To the best of our knowledge, this is the first work for IP protection of multi-exit models by fingerprinting. 
(2) We use a novel idea by fingerprinting DNN models via inference time which is more effective than all prior existing fingerprinting methods. 
(3) We conduct extensive experiments (totally $\sim1000$ models trained with $\sim 2000$ GPU hours) on 3 datasets (CIFAR-10/100, Tiny-ImageNet) with 3 mainstream architectures (ResNet-56, VGG-16, and MobileNet) to prove the \textit{uniqueness} and \textit{robustness} of our method.

\vspace{-1em}
\section{Background}\vspace{-1em}

\subsection{Multi-exit DNN Models}
\vspace{-1ex}
The increasing performance of DL tasks today brings an increasing number of layers in most of the state-of-the-art DNN models. 
However, such increasing complex models generate excessively complex representations which lead to the overthinking issue. 
On the one hand, forcing canonical samples to inference all layers of a DNN model definitely brings a waste of energy and time~\citep{huang2017multi}. 
On the other hand, \cite{kaya2019shallow} pointed out that a further computation on a simple sample at the deeper layers will lead to misclassification. 
Moreover, since many DNN models are supposed to be deployed on resource-aware or resource-constrained scenarios such as the Internet of Things (IoT), complex DNN models are inefficient or impractical to be used. 

One promising approach to solve these issues is to introduce the input-adaptive DNNs. 
There are two main types of input-adaptive DNNs: adaptive neural networks (AdNNs)~\citep{wang2018skipnet, figurnov2017spatially} which are only applicable for ResNet-like models; the multi-exit DNNs~\citep{teerapittayanon2016branchynet, huang2017multi, kaya2019shallow} that are more generic and widely used. 
Specifically, a multi-exit DNN introduces multiple exit points on a vanilla DNN to allow the inference to preemptively finish at one of the exit points when the network is sufficiently confident with a pre-defined stop criteria. 
By letting most of the input samples exit at an earlier layer of a complex DNN model, such multi-exit structures are found much more efficient compared with previous works like DNN split learning task \citep{hauswald2014hybrid,kang2017neurosurgeon,eshratifar2019jointdnn,he2019model,he2020attacking} which every sample has to go through all the distributed layers for prediction. 
Early multi-exit structures are designed specifically such as MSDNets~\citep{huang2017multi} which cannot leverage the off-the-shelf DNN models. 
Later on, shallow-deep network (SDN)~\citep{kaya2019shallow} was proposed as a generic structure to transform any off-the-shelf model into a multi-exit model which is more effective and efficient. 
Note that in this paper we only experiment with SDN structure and the other multi-exit structures can be experimented in a similar way.

\subsection{DNN IP Protection}
\vspace{-1ex}

Training large DNN models are becoming more and more costly. 
For instance, training a state-of-the-art GPT-3 model with 175 billion parameters requires millions of dollars~\citep{GPT3_cost}. 
Therefore, DNN models are becoming valuable and important intellectual property (IP).
However, recent research has shown that the IP of DNN models is vulnerable to illegal copy, redistribution, abuse~\citep{xue2021dnn} or model stealing attacks~\citep{zhu2021hermes}. 
Specifically, an attacker may get a stolen or illegally copied DNN model and deploy it on its own server for illegal profit. 
This brings a new challenge for the model owners to verify the ownership of suspicious models in a black-box scenario. 
The early approach to solve this challenge is to watermark a DNN model relying on using out-of-distribution features~\citep{adi2018turning, zhang2018protecting} or handcrafted parameters~\citep{tang2020embarrassingly}. 
This watermarked model will behave normally on benign samples but output specific labels for watermarked samples to verify the model's IP. 
The limitations of this watermarking approach are clear. 
First, the watermarking approach will decrease the model accuracy which is unacceptable in critical applications~\citep{cao2021ipguard}. 
Second, watermarking procedures are mostly combined with the training phase which cannot be used for off-the-shelf models. 
Third, once the watermark of one DNN model is leaked, the adversary can easily bypass this verification adaptively but it will be too costly for the model owner to rebuild another model with a different watermark.

\begin{wrapfigure}[13]{r}{0.46\textwidth}
    \centering
    \vspace{-3ex}
    \includegraphics[width=0.46\textwidth]{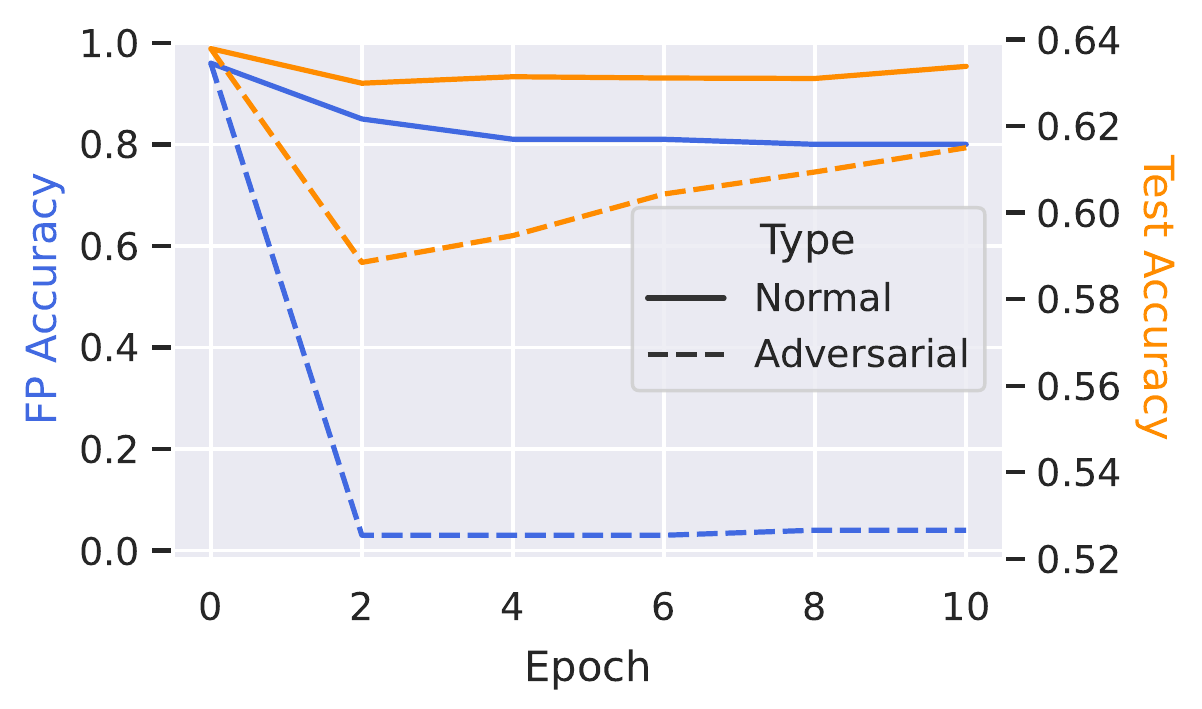}
    \vspace{-5ex}
    \caption{Adversarial training mitigates fingerprint \citep{conferrable} in a few epochs.}
    \label{fig:adversarial_finetuning}
\end{wrapfigure}

The more promising approach to verify the model ownership was proposed as the DNN fingerprinting. 
The general idea of the DNN fingerprinting method is to verify the model ownership via its 
\textit{unique and robust} feature.
However, existing DNN fingerprinting methods~\citep{cao2021ipguard, conferrable, characteristic_example, wang2021fingerprinting} mainly rely on the inference output of special samples generated in a similar way as AEs~\citep{goodfellow2014explaining}. 
For instance, they generate fingerprint samples to mislead model prediction to certain labels for IP verification by releasing the perturbation bounds~\citep{cao2021ipguard} or limiting the transferability~\citep{conferrable}. 
The first limitation of these AE-based fingerprinting is that they are vulnerable to model robustness enhancement such as adversarial training (see the example we reproduced in \figurename~\ref{fig:adversarial_finetuning}, experimentation details in Appendix~\ref{appendix:limitation_aebased}).
The second limitation to use them for fingerprinting the multi-exit model is that the prediction accuracy for both benign and fingerprint samples is easily manipulated by the tunable early-exit criteria of multi-exit models (decrease exit criteria will lead to lower model performance but faster inference on average).  
Therefore, we argue that the existing DNN fingerprinting methods for multi-exit models are impractical since the adversary can manipulate the prediction results by (1) retraining the backbone DNN model via adversarial training or (2) significantly changing the multi-exit structure to decrease the overall accuracy. 

\vspace{-1ex}
\section{Methodology}
\vspace{-1em}

\subsection{Threat Model} 
\vspace{-1ex}

The \textbf{adversary's goal} is to deploy an illegal copy or stolen multi-exit model in a resource-constrained scenario for illegal profit and disable the model ownership verification. Note that if the adversary wishes to use only the trained backbone model, there is no need to steal a multi-exit model and extract the backbone model only. 
  
The \textbf{adversary's knowledge} is that the target multi-exit model and a partial training dataset is known to the adversary. 
Specifically, the adversary knows all the model details including the backbone model's structure, parameters, and the ICs. 
Also, we argue the necessity to consider that an adversary may also have a partial training dataset (at lease the dataset of the same distribution) as well. 

The \textbf{adversary's capability} is that the adversary has the capability to modify the model but cannot train a new model. 
We assume that the attacker will try to modify the target model to avoid the IP verification but try to keep the model performance.
The practical modification includes modifications on the backbone model (\eg finetuning, compression, pruning, or adversarial training) or modification of the multi-exit structure (\eg removing, modifying, adding ICs, or changing early exit criteria).

\vspace{-1ex}
\subsection{Methodology Overview}
\vspace{-1ex}

The idea of our methodology is to verify the fingerprint samples via inference but do not rely on their predicted labels. 
Practically, for the multi-exit models, we verify the IP of the target model by comparing the inference time of specialized fingerprint samples with a pre-defined threshold. 
Since the initial motivation and later designs of the multi-exit model is to encourage as many samples as possible to exit earlier to avoid overthinking and to improve efficiency, our fingerprint samples are designed to behave significantly differently in this exit scenario. 
Our fingerprint samples will be dedicated to exit as late as possible which will consume a significantly longer time for inference than benign samples on average. 
This special inference time is the natural feature of this particular model for us to verify the IP. 
On the one hand, since our method does not care about the predicted labels for the fingerprint samples, those modifications on manipulating accuracy or model enhancement are not effective to mitigate our fingerprint samples. 
This can achieve the \textit{robustness} of our method.   
On the other hand, since our fingerprint samples are significantly related to the target model's parameters, we can effectively distinguish our target model from the independent models which are trained with different initial parameters.
Thus, our method can achieve \textit{uniqueness}. 

\vspace{-1ex}
\subsection{Fingerprinting Method}
\label{method}
\vspace{-1ex}

\noindent\textbf{Notations.} 
Let $n$ denote the number of exits (including exits of internal classifiers (IC) and the last-layer exit point) of the SDN model and $\theta_i$ denote the model parameters up to the $i$-th exit, where $1\leq i\leq n$ and $\theta_1\subset\dots\subset\theta_i\subset\theta_{i+1}\dots\subset\theta_n$. 
We denote the softmax probability output (\ie confidence vector) $f_{\theta_i}(x)$ given the input $x$ at the $i$-th exit as $f_i(x)$. 
For setting the early exit criteria, let  $T_c$ denote a confidence-based exit threshold, such that the inference of input $x$ stops at the $i$-th exit iff $\max(f_i(x))\geq T_c$ and $\forall j<i, \max(f_j(x))<T_c$.

\noindent\textbf{Fingerprint generation.}
Given a benign input sample $x$, we aim to craft a perturbation $\delta_x$ to manipulate the modified input sample $x+\delta_x$ to not exit at any early exit point. 
Note that this generation process is not related to any specific or expected predicted label of the modified sample which is significantly different from the idea of adversarial attacks. 
Then, we can use the inference time of a set of such modified samples to fingerprint the target model. 
Formally, we define the following optimization loss to craft $\delta_x$ for $x$:

\begin{equation}
    \label{eq:fingerprinting_loss}
    L_{f}(x) = \sum\limits_{i=1}^{n-1}D_\mathrm{KL} (f_i(x), u) - D_\mathrm{KL} (f_n(x), u)
\end{equation}

where $D_\mathrm{KL}$ is the Kullback-Leibler (KL) divergence which can be seen as the distance between two distributions and $u$ is the uniform vector of the same length as the confidence vector, \ie $u=[\frac{1}{n_y}]_{1\leq i\leq n_y}$. 

The first term of the summation in Equation~(\ref{eq:fingerprinting_loss}) pushes the confidence vectors of internal exits towards a uniform distribution
On the contrary, the last term pushes the confidence vector of the final layer away from the uniform distribution. 
Such that the confidence scores of the ICs will be lower than the exit threshold $T_c$ and will not exit. 
As a result, these fingerprint samples will not be able to exit at any early exit point and the inference time of them will be significantly longer than benign samples on average. 
Particularly, since our fingerprint generation does not rely on the samples' predicated labels, those model enhancement methods that mitigate perturbations for misleading (\eg adversarial training) are not effective anymore. 
Thus, we claim the fingerprinting \textit{robustness} is improved compared with existing fingerprinting methods (see experimentation in Section 4.3).

We adopt the $L_2$-based CW optimization technique to improve the fingerprinting \textit{uniqueness} by reducing the fingerprints' transferability to independent models. 
Typically, suppose the selected benign sample is $x$ and the fingerprint to craft is $x'$. Then the final loss is
\begin{equation}
    \label{eq:cw_optimize}
    L(x') = \norm{x-x'}_2 + cL_f(x'),
\end{equation}
where $c$ is a parameter used to balance the stealthiness and the effectiveness of $x'$. 
A larger $c$ leads to better fingerprinting uniqueness, at the expense of fingerprints' visual similarity to the original samples. 
Note that our method does not set any perturbation bounds on such fingerprint sample generation. 
However, the experimentation results indicate that with our chosen parameter $c$, the visual similarity of our fingerprint samples is very similar to AEs which are also human imperceptible (see statistical results in Section 4.4 and visual results in Appendix~\ref{appendix:additional}).

\noindent\textbf{Fingerprint Verification.}
The model owner first generates $N$ fingerprint samples using the method described above.
Then, the model owner feeds the $N$ fingerprint samples to the suspicious model for inference and records the inference time of these samples. 
Note that a longer inference time indicates a deeper exit layer. 
Practically, in order to get a stable inference time, we calculate the inference time by averaging the time of each sample for 10 times.

\begin{wrapfigure}[14]{r}{0.46\textwidth}
    \centering
    \vspace{-3ex}
    \includegraphics[width=0.46\textwidth]{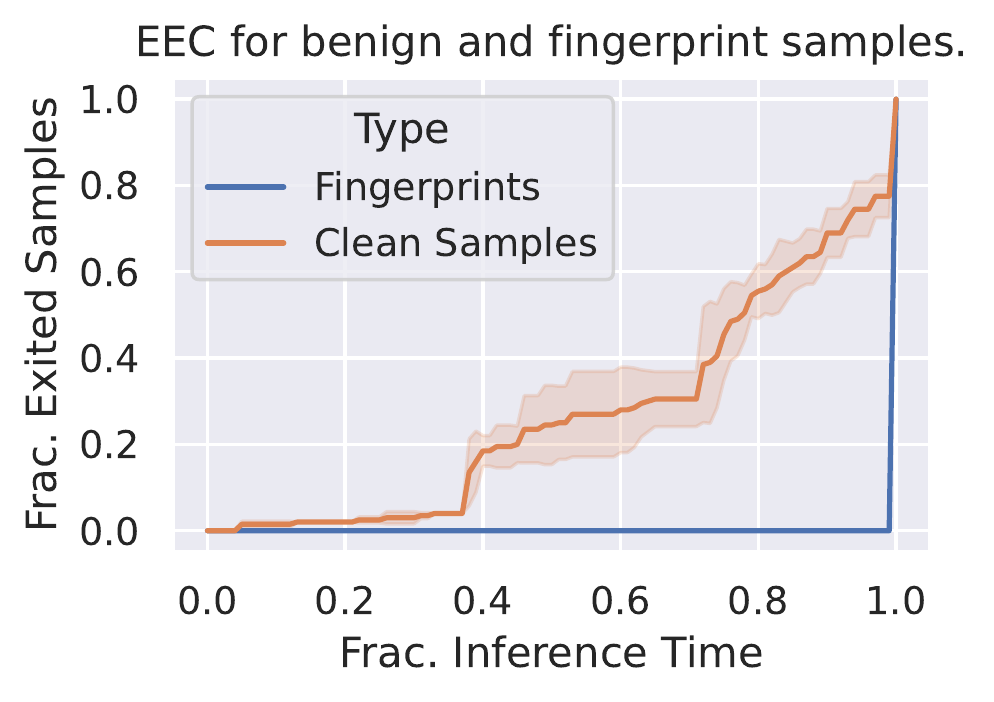}
    \vspace{-6ex}
    \caption{EEC of fingerprint and benign samples (average of 10 different groups).}
   \label{fig:eec_clean_samples_fingerprints}
\end{wrapfigure}

In order to get a more reliable fingerprinting verification, we use an EEC AUC score rather than directly using the inference time for calculation. 
The definition is explained with an example as follows. 
First, to quantify the suspicious model's inference time on $N$ fingerprint samples, we modified the early-exit capacity (EEC) curve~\citep{hong2020panda} to measure the inference time costs for these $N$ fingerprint samples. 
An EEC used in~\citep{hong2020panda} describes how a set of samples exit a multi-exit model by calculating the ratio of samples exiting at each point.
Here we modify the EEC by measuring the exit situation by inference time. 
For instance, we feed 10 groups of 100 benign samples to a multi-exit model and record the inference time.
All inference time is normalized within [0,1] by comparing with the longest inference time. 
Here we assume the longest inference time as the maximum inference time of these benign samples since there will always be benign samples exit at the last layer. 
The results of the 10 groups and their average value are in Figure~\ref{fig:eec_clean_samples_fingerprints}. 
For benign samples, 60\% of them can exit within 0.8 of the longest inference time. 
However, if we use 100 fingerprint samples, 100\% of them will pass all the layers so the EEC is significantly different. 

Second, we use the area under the curve (AUC) of the EEC of the generated $N$ fingerprint samples to determine whether an IP verification is successful or not. 
Specifically, we denote the AUC of the EEC for $N$ samples as $T_N$. 
For fingerprint samples, the corresponding $T_N$ is significantly lower than the benign samples. 
Thus, for the IP verification, we compared $T_N$ of the fingerprint samples inference by the suspicious model with a pre-defined threshold $T_f$. 
If $T_N < T_f$, we claim the ownership verification is a success and the suspicious model compromises the IP of the target model. 
Otherwise, the suspicious model is an independent model. 

\vspace{-1ex}
\section{Experimentation}
\vspace{-1em}

In this section, we evaluate our fingerprinting method from two aspects including \textit{robustness} and \textit{uniqueness}. 
Specifically, we compare our method with the baseline~\citep{conferrable} to show that: (1) we achieve better uniqueness with more accurate model verification, and (2) we are robust to adversarial training which can easily mitigate this AE-based baseline fingerprinting method.

\vspace{-1ex}
\subsection{Experimental Setup} 
\label{subsec:exp_setup}
\vspace{-1ex}

\noindent\textbf{Datasets \& Model Structure.}
We use the image classification datasets CIFAR-10/100 \citep{CIFAR10}, noted as ``C10'' and ``C100'' respectively and 200-class
Tiny-ImageNet\footnote{\url{https://www.kaggle.com/c/tiny-imagenet}} noted as ``TI''. 
For the model structure, we choose three structures including ResNet-56~\citep{he2016deep}, VGG-16~\citep{vgg16}, and MobileNet~\citep{howard2017mobilenets}. 
For each structure, we train models on C10 and C100 for 50 epochs and on TI for 100 epochs. 
For the multi-exit model structure, we use the SDN structure~\citep{kaya2019shallow} and other structures can be fingerprinted in a similar way. 
We transform the trained models into SDN models by adding ICs at some hidden layers and train the ICs for 25 epochs. 
In the end, we have the multi-exit ResNet-56 model with 27 ICs, the multi-exit VGG-16 model with 14 ICs, and the multi-exit MobileNet model with 14 ICs.

\noindent\textbf{Independent Model Training.}
For evaluating the \textit{uniqueness}, we train independent models for fingerprinting verification. 
The independent models refer to the models that use the exact same training dataset, structure, hyper-parameters but only with different initial parameters. 
These independent models are similar to the target model but should be excluded from IP verification. 

For each dataset and each architecture, we train 100 models from scratch with different initialization and random seeds and transform them into SDNs as independent models. 
Thus, in this paper, we trained 900 models in total from 3 different datasets and 3 different model structures\footnote{Note here the training for these 900 independent models takes about 1,600 GPU hours.}. 
Before experimenting with fingerprinting on one target model, we report the averaged EEC AUC scores on 100 benign samples across the 100 models for each structure and each dataset in Table~\ref{tab:eec_score} of Appendix~\ref{appendix:additional} (the standard deviations are less than 0.02). 
Remark that the ResNet-56 model trained on CIFAR-10 has the lowest clean EEC AUC score, because $T_c$ for RAD-5 is close to 1 (\ie, 0.95), and there will be fewer early exits during the inference.
Training details can be found in Appendix~\ref{appendix:details}.

\noindent\textbf{Adversarial Modifications.} 
For proving \textit{robustness}, we consider comprehensive model modifications that an adversary can make with two categories.  
The first one is to modify the multi-exit model structure by manipulating the ICs. 
Since training ICs requires significantly less resources than training DNNs from scratch, it is reasonable for an adversary to modify ICs to avoid IP verification. 
The adversary may (1) remove ICs or add ICs to change the total number of exit points which will influence samples' overall prediction accuracy. 
Since our fingerprint samples suppose to exit at the last layer of the model, removing or adding ICs can not affect our verification. 
Also, they can (2) retrain only the ICs to influence the prediction accuracy. 
Moreover, they can (3) change the early exit criteria for decreasing inference accuracy to compromise the fingerprint samples' prediction labels. 
For instance, the early-exit criteria for a multi-exit model could refer to a confidence threshold $T_c$. 
A lower $T_c$ encourages more samples to exit at earlier layers to speed up the inference but decreases prediction accuracy noted as of relative accuracy drop (RAD). 
Note that a significant decrease in prediction accuracy may disable fingerprinting but makes no sense since such a model is useless. 
We assume the maximum RAD that can be accepted is 15\% so we choose two thresholds (RAD as 5\% and 15\%) for our experimentation which are denoted as RAD=5 and RAD=15 respectively.

The second kind of adversarial modification is on the extracted backbone DNN model. 
Here we consider a comprehensive scenario that includes compression, pruning, finetuning, and adversarial training. 
We perform the model pruning with rates of 0.1, 0.2, 0.3, and 0.4 to get pruned models and compress with 8-bit to get quantized models. 
We apply finetuning for 16 epochs for all target models.
For adversarial training, we retrain 6 epochs for C10 and 10 epochs for C100 and TI. 
We save the modified model every 2 epochs. 
To avoid randomness during finetuning, we finetune the model 10 times independently under the same setting. 
In total, there are 80 finetuned models for each target model, 30 adversarial trained models for each target model on C10, and 50 adversarial trained models for each target model on C100 and TI respectively. 
We then transform them back into SDN multi-exit models.
Experiment details of finetuning and adversarial training are in Appendix~\ref{appendix:details}.

\noindent\textbf{Metrics.}
We use three metrics to evaluate the effectiveness of our method. 
The first is $T_N$ (EEC AUC score) for the $N$ fingerprint samples on a suspicious model. 
Since $T_N$ is directly used for the IP verification by comparing with the pre-defined threshold $T_f$, a value of $T_N$ smaller than $T_f$ for a suspicious model refers to a success of IP verification. 
For evaluating fingerprinting of multiple suspicious models, we calculate the IP verified rate as our second metric to indicate the ratio of models verified as stolen models. 

The third metric is the receiver operating characteristic (ROC) with AUC~\citep{conferrable} to compare our method with the baseline. 
The ROC curve consists of multiple pairs of the false positive rate (FPR) and the true positive rate (TPR) computed for various detection thresholds. 
Since IP verification is a bi-class problem (IP verified or not), we use the AUC score of ROC for a straightforward comparison. 
For instance, if a classifier is perfect, the ROC AUC should be equal to 1. 

\noindent\textbf{Baseline.} 
We choose a state-of-the-art AE-based fingerprinting~\citep{conferrable} as the baseline. 
Specifically, we generate fingerprint samples by reproducing their method to craft fingerprint samples and set the expected prediction labels. 
When verifying the IP, the model owner sends the crafted fingerprint samples to the suspicious model and gets the returned labels. 
Then, the model owner calculates the proportion of how many returned labels are the same as his expected labels, and if it is higher than a threshold, the model owner can claim the IP is compromised.

\begin{wraptable}[8]{r}{0.5\textwidth}
\vspace{-1em}
\centering
\caption{Pre-defined thresholds $T_f$ in this paper.}
\label{tab:selected_th}
\resizebox{0.99\linewidth}{!}{
\begin{tabular}{c|c|c|c|c}
\Xhline{1pt}
\textbf{RAD} & \textbf{Dataset} & \textbf{ResNet-56} & \textbf{VGG-16} & \textbf{MobileNet} \\ \Xhline{1pt}
\multirow{3}{*}{RAD=5} & C10 & 0.006 & 0.204 & 0.388 \\ \cline{2-5} 
 & C100 & 0.023 & 0.040 & 0.403 \\ \cline{2-5} 
 & TI & 0.011 & 0.013 & 0.097 \\ \Xhline{1pt}
\multirow{3}{*}{RAD=15} & C10 & 0.076 & 0.449 & 0.644 \\ \cline{2-5} 
 & C100 & 0.130 & 0.176 & 0.566 \\ \cline{2-5} 
 & TI & 0.063 & 0.024 & 0.244 \\ \Xhline{1pt}
\end{tabular}}
\end{wraptable}

\noindent\textbf{Threshold Setting.}
To compare $T_N$ (EEC AUC score) for the $N$ fingerprint samples with a pre-defined threshold $T_f$, a proper $T_f$ must be set. 
Since the $T_N$ of fingerprint samples (normally close to 0) is significantly less than benign samples, a higher $T_f$ leads to more success rate of IP verification. 
However, increasing the $T_f$ will lead to potential false positives on misclassifying independent models. 
Based on such a trade-off, we list the pre-defined $T_f$ in Table~\ref{tab:selected_th} for our experimentation in this paper. 
Also, an ablation study on setting $T_f$ is in Section 4.4, Table~\ref{tab:th_accs_abaltion}.

\vspace{-1ex}
\subsection{Fingerprint Uniqueness: Results and Analysis}
\vspace{-1ex}

To prove the \textit{uniqueness} of our method, we show two kinds of experimentation results. 
The first is to prove that our method can accurately detect our model from various independent models.
The second is to compare our method with the baseline on the same detection task.

\begin{wraptable}[7]{r}{0.68\textwidth}
\vspace{-4ex}
\centering
\caption{IP verified rate: 100 independent and 100 shadow models.}
\label{tab:acc_selected_th}
\resizebox{0.99\linewidth}{!}{
\begin{tabular}{c|c|c|c|c|c|c|c|c|c|c}
\Xhline{1pt}
\multirow{2}{*}{\textbf{RAD}} & \multirow{2}{*}{\textbf{}} & \multicolumn{3}{c|}{\textbf{ResNet-56}} & \multicolumn{3}{c|}{\textbf{VGG-16}} & \multicolumn{3}{c}{\textbf{MobileNet}} \\ \cline{3-11} 
 &  & C10 & C100 & TI & C10 & C100 & TI & C10 & C100 & TI \\ \Xhline{1pt}
\multirow{2}{*}{RAD=5} & Indep. & 0.50 & 0.07 & 0.01 & 0.02 & 0.03 & 0.00 & 0.02 & 0.00 & 0.02 \\ \cline{2-11} 
 & Stolen & 0.60 & 0.93 & 0.98 & 0.98 & 0.97 & 0.99 & 0.98 & 1.00 & 0.94 \\ \Xhline{1pt}
\multirow{2}{*}{RAD=15} & Indep. & 0.00 & 0.01 & 0.03 & 0.01 & 0.02 & 0.06 & 0.01 & 0.00 & 0.01 \\ \cline{2-11} 
 & Stolen & 0.82 & 0.98 & 0.99 & 0.99 & 0.98 & 0.94 & 0.99 & 1.00 & 0.98 \\ \Xhline{1pt}
\end{tabular}}
\end{wraptable}

We first randomly choose 100 models from the modified models (\eg compression, pruning, finetune, adversarial training) as the stolen models. 
Then, we use the trained 100 independent models (only initial parameters are different) for evaluating our fingerprinting method. 
Results in Table~\ref{tab:acc_selected_th} show that our method can effectively verify the IP of the plagiarized models (\eg 0.98 for TI dataset, ResNet-56 with RAD=5) and keep a low misdetection ratio on independent models (almost 0 IP verified for most cases).

\begin{wraptable}[7]{r}{0.68\textwidth}
\vspace{-4ex}
\centering
\caption{ROC AUC scores for our fingerprinting and baseline.  
}
\label{tab:roc_auc_independent}
\resizebox{0.99\linewidth}{!}{
\begin{tabular}{c|c|c|c|c|c|c|c|c|c|c}
\Xhline{1pt}
\multirow{2}{*}{\textbf{RAD}} & \multirow{2}{*}{\textbf{Method}} & \multicolumn{3}{c|}{\textbf{ResNet-56}} & \multicolumn{3}{c|}{\textbf{VGG-16}} & \multicolumn{3}{c}{\textbf{MobileNet}} \\ \cline{3-11} 
 &  & C10 & C100 & TI & C10 & C100 & TI & C10 & C100 & TI \\ \Xhline{1pt}
\multirow{2}{*}{RAD=5} & Baseline & \textbf{0.72} & 0.84 & 0.94 & 0.75 & 0.94 & 0.92 & 0.97 & 0.94 & 0.91 \\ \cline{2-11} 
 & Ours & 0.56 & \textbf{0.99} & \textbf{1.00} & \textbf{0.99} & \textbf{0.99} & \textbf{1.00} & \textbf{1.00} & \textbf{0.99} & \textbf{1.00} \\ \Xhline{1pt}
\multirow{2}{*}{RAD=15} & Baseline & 0.73 & 0.82 & 0.93 & 0.78 & 0.90 & 0.92 & 0.64 & 0.94 & 0.98 \\ \cline{2-11} 
 & Ours & \textbf{0.90} & \textbf{1.00} & \textbf{1.00} & \textbf{0.99} & \textbf{1.00} & \textbf{1.00} & \textbf{0.99} & \textbf{1.00} & \textbf{1.00} \\ \Xhline{1pt}
\end{tabular}}
\end{wraptable}

Second, we compare our method with the baseline \citep{conferrable} in Table~\ref{tab:acc_selected_th}. 
Here we use the ROC AUC score to show the detection result. 
Note that if we see IP verification as a bi-class problem, a good classifier should have a ROC AUC score close to 1. 
Table~\ref{tab:roc_auc_independent} shows that our method has clearly better ROC AUC scores on almost all cases. 
For MobileNet on C10, the ROC AUC score of baseline drops from 0.97 to 0.64 when the RAD changes from 5 to 15, indicating that manipulating multi-exit model ACC can mitigate AE-based fingerprinting methods.
In summary, we claim that our method achieves uniqueness and outperforms the baseline.

\vspace{-1ex}
\subsection{Fingerprint Robustness: Results and Analysis}
\vspace{-1ex}

For proving the \textit{robustness}, we first provide the results of our method for all adversarial modification cases\footnote{Using multiple model modification on one target model is unusual but possible, see results in Appendix~\ref{appendix:additional}.}. 
The metric we used here is to give the EEC AUC scores for our 100 fingerprint samples. 
Note they are all significantly less than the threshold $T_f$. 
Then, we compare our method with the baseline on adversarial training only since the baseline can achieve robustness for the other modifications.

\begin{wraptable}[6]{r}{0.68\textwidth}
\vspace{-2em}
\centering
\caption{Average EEC AUC for 100 models of IC retraining.}
\label{tab:icretrain_eec}
\resizebox{0.99\linewidth}{!}{
\begin{tabular}{c|c|c|c|c|c|c|c|c|c}
\Xhline{1pt}
\multirow{2}{*}{\textbf{RAD}} & \multicolumn{3}{c|}{\textbf{ResNet-56}} & \multicolumn{3}{c|}{\textbf{VGG-16}} & \multicolumn{3}{c}{\textbf{MobileNet}} \\ \cline{2-10} 
 & C10 & C100 & TI & C10 & C100 & TI & C10 & C100 & TI \\ \Xhline{1pt}
RAD=5 & 0.01 & 0.01 & 0.01 & 0.01 & 0.01 & 0.01 & 0.02 & 0.01 & 0.03 \\ \hline
RAD=15 & 0.01 & 0.01 & 0.01 & 0.01 & 0.01 & 0.02 & 0.14 & 0.10 & 0.05 \\ \Xhline{1pt}
\end{tabular}}
\end{wraptable}

\noindent\textbf{Model IC retraining.}
We apply IC retraining to the owner's multi-exit model (SDN structure). 
We then generate 100 stolen models with the same backbone DNN models but only differ in ICs. 
Note that our fingerprint samples are supposed to exit at the last layer so adding or removing ICs has no influence on our method. 
We list the EEC AUC scores for these 100 modified models. 
It can be observed that these scores are pretty close to the ideal value 0 and are significantly less than the threshold $T_f$. 
Thus, the IP of these 100 plagiarized models can all be successfully verified by our method.

\begin{wraptable}[6]{r}{0.68\textwidth}
\vspace{-2ex}
\centering
\caption{EEC AUC scores for quantized target models.}
\label{tab:quant_eec}
\resizebox{0.99\linewidth}{!}{
\begin{tabular}{c|c|c|c|c|c|c|c|c|c}
\Xhline{1pt}
\multirow{2}{*}{\textbf{RAD}} & \multicolumn{3}{c|}{\textbf{ResNet-56}} & \multicolumn{3}{c|}{\textbf{VGG-16}} & \multicolumn{3}{c}{\textbf{MobileNet}} \\ \cline{2-10} 
 & C10 & C100 & TI & C10 & C100 & TI & C10 & C100 & TI \\ \hline
RAD=5 & 0.00 & 0.00 & 0.00 & 0.00 & 0.00 & 0.00 & 0.00 & 0.02 & 0.06 \\ \hline
RAD=15 & 0.00 & 0.00 & 0.00 & 0.00 & 0.00 & 0.00 & 0.01 & 0.04 & 0.06 \\ \Xhline{1pt}
\end{tabular}
}
\end{wraptable}

\noindent\textbf{Model compression.}
we plot the EEC AUC scores for detecting suspicious models that are generated by compressing the target models. 
We can see that the compression has little effect on the EEC AUC scores since most of them stay around 0. 
According to our threshold $T_f$, the IP of all these suspicious models can be successfully verified. 
Thus, we claim that the model compression has no influence on our fingerprinting method.

\begin{figure}[htbp]
    \vspace{-2ex}
    \centering
    \includegraphics[width=\textwidth]{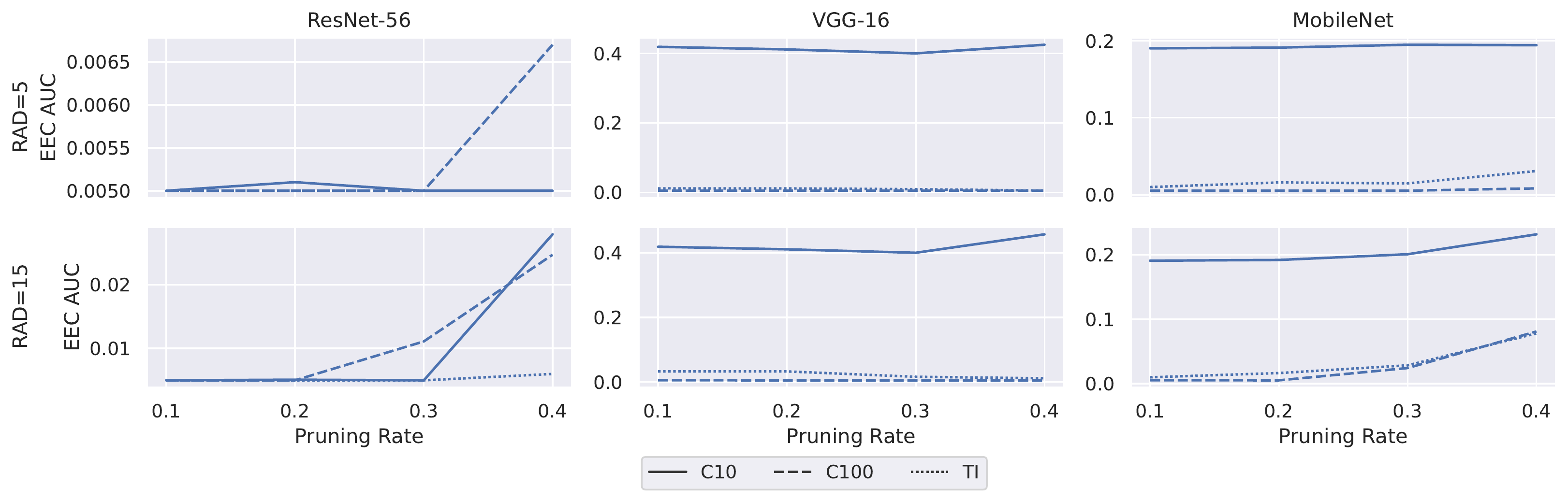}
    \vspace{-5ex}
    \caption{EEC AUC scores for different pruning rates.
    }
    \label{fig:allsingleprune}
\end{figure}

\noindent\textbf{Model pruning.} 
We set the pruning rate from 0.1 to 0.4 to get suspicious models. 
A large pruning rate introduces significant ACC loss (pruning rate of 0.4 decreases ACC by about 15\%) so a larger pruning rate will make the model useless. 
The EEC AUC scores of the pruning case are in Figure~\ref{fig:allsingleprune}.  
We find that different models have different sensitivity for pruning. 
But all EEC AUC scores stay lower than the threshold so still, all IP verification will succeed on these suspicious models.

\begin{figure}[!htbp]
    \vspace{-2ex}
    \centering
    \includegraphics[width=\textwidth]{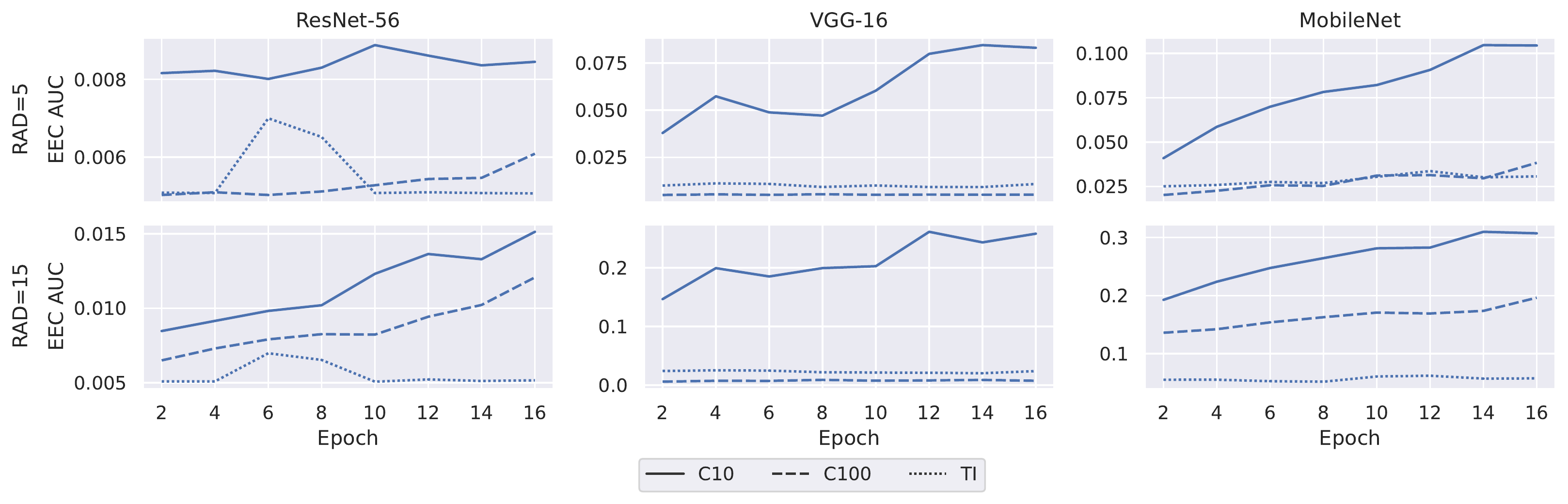}
    \vspace{-5ex}
    \caption{EEC AUC scores for different finetuning epochs.}
    \label{fig:allfinetune_RAD5}
\end{figure}

\noindent\textbf{Model finetuning.} 
For evaluating the model finetuning case, we use benign samples for the finetuning. 
We calculate the EEC AUC score for the 80 finetuned models generated from each target model and list the average EEC AUC score in Figure~\ref{fig:allfinetune_RAD5} (with tiny standard deviations). 
It can be observed the model finetuning has little influence on the EEC AUC scores as well. 
The IP verification can succeed on all model finetuning cases by our fingerprinting method. 

\noindent\textbf{Model adversarial training.}
We compare our method with the baseline for the adversarial training case. 
Here we use the IP verified rate to evaluate both methods. 
Experiment details on adversarial training are in Appendix~\ref{appendix:details}. 
In Figure~\ref{fig:advfinetune_RAD5}, we list the comparison results for RAD=5 (the left three columns) and for RAD=15 (the right three columns). 
Note that for RAD=15, the baseline directly fails for C10 MobileNet and C100 MobileNet since the prediction accuracy is decreased which also changes the fingerprint samples' prediction labels.
Our method is not sensitive to adversarial training in most cases while the baseline method fails in most cases after even a few epochs. 
The main reason is that our method is not verified according to the prediction labels by suspicious models such that enhancing model robustness by adversarial training has no effects. 

\begin{figure}[!htbp]
    \centering
    \includegraphics[width=\textwidth]{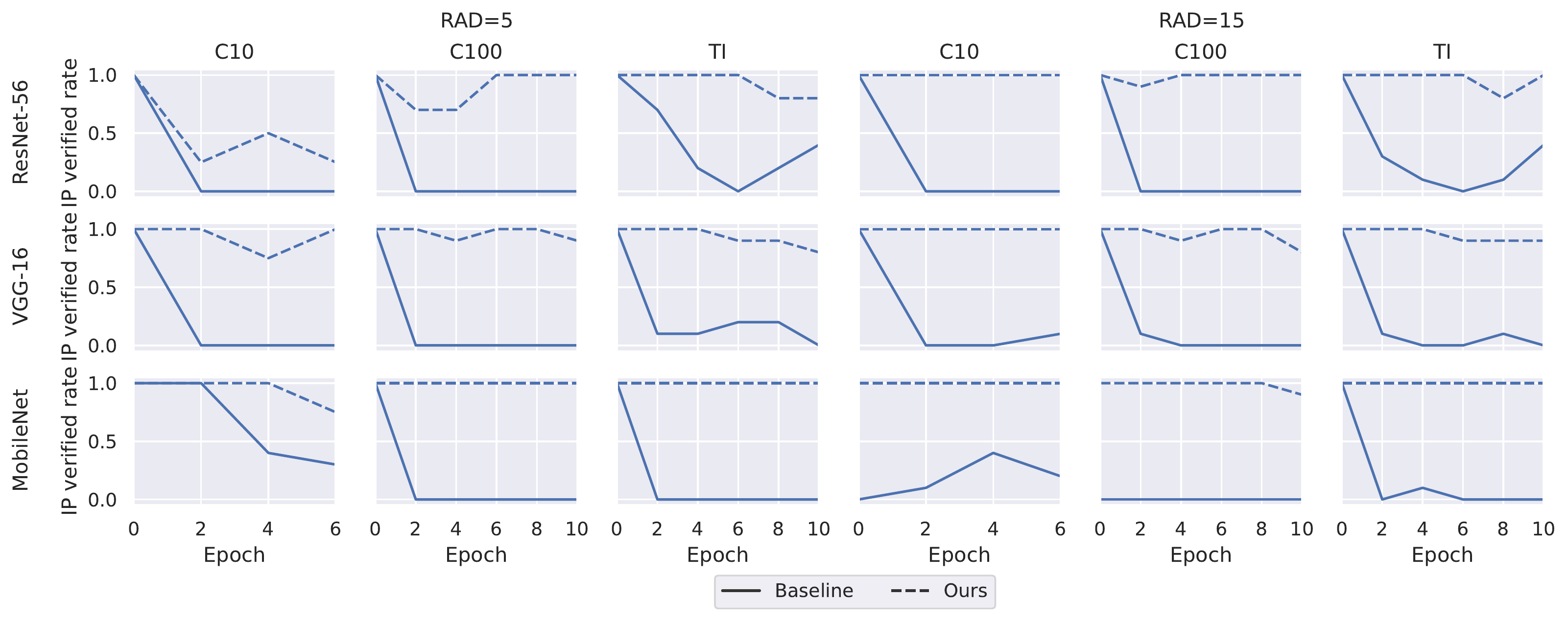}
    \vspace{-5ex}
    \caption{IP verified rate of models modified by adversarial training under RAD=5 and RAD=15.}
    \label{fig:advfinetune_RAD5}
\end{figure}

\subsection{Ablation Study}

We give the visual difference of our fingerprint samples and the results for different thresholds $T_f$.

\begin{wraptable}[5]{r}{0.68\textwidth}
\vspace{-2em}
\centering
\caption{L2 score for our method and baseline.}
\label{tab:visual_similarity}
\resizebox{0.99\linewidth}{!}{
\begin{tabular}{c|c|c|c|c|c|c|c|c|c}
\Xhline{1pt}
\multirow{2}{*}{\textbf{Method}} & \multicolumn{3}{c|}{\textbf{ResNet-56}} & \multicolumn{3}{c|}{\textbf{VGG-16}} & \multicolumn{3}{c}{\textbf{MobileNet}} \\ \cline{2-10} 
 & C10 & C100 & TI & C10 & C100 & TI & C10 & C100 & TI \\ \Xhline{1pt}
Baseline & 0.005 & 0.014 & 0.009 & 0.005 & 0.020 & 0.044 & 0.005 & 0.010 & 0.013 \\ \hline
Ours & 0.019 & 0.008 & 0.023 & 0.033 & 0.056 & 0.034 & 0.005 & 0.009 & 0.018 \\ \Xhline{1pt}
\end{tabular}}
\end{wraptable}

\noindent\textbf{Visual similarity.}
The L2 scores averaged over 100 samples of baseline and fingerprint samples are in Table~\ref{tab:visual_similarity}. 
We can see that the similarity of our fingerprint samples to the original samples is at the same level as the baseline. 
Most of the L2 scores of our fingerprint samples are less than 0.05 which are visually imperceptible to human eyes according to the definition of AE~\citep{carlini2017towards}. 
We refer readers to Appendix~\ref{appendix:details} for more details of generation fingerprint samples and Appendix~\ref{appendix:additional} visual results.

\noindent\textbf{Threshold $T_f$.}
Setting different thresholds $T_f$ for the fingerprint verification will either decrease the fingerprint accuracy for independent models or the shadow model. 
We tune the threshold $T_f$ and calculate the IP verified rate for the same task in Section 4.2 with the results in Table~\ref{tab:th_accs_abaltion}. 

\begin{table}[!h]
\centering
\vspace{-3ex}
\caption{IP verified rate: 100 independent models and 100 shadow models for different $T_f$.}
\label{tab:th_accs_abaltion}
\resizebox{0.8\linewidth}{!}{
\begin{tabular}{c|c|c|c|c|c|c|c|c|c|c}
\Xhline{1pt}
\multirow{3}{*}{\textbf{\begin{tabular}[c]{@{}c@{}}RAD\end{tabular}}} & \multirow{3}{*}{\textbf{Dataset}} & \multicolumn{3}{c|}{\textbf{ResNet-56}} & \multicolumn{3}{c|}{\textbf{VGG-16}} & \multicolumn{3}{c}{\textbf{MobileNet}} \\ \cline{3-11} 
 &  & $T_f$ & \begin{tabular}[c]{@{}c@{}}Indep.\end{tabular} & \begin{tabular}[c]{@{}c@{}}Shad.\end{tabular} & $T_f$ & \begin{tabular}[c]{@{}c@{}}Indep.\end{tabular} & \begin{tabular}[c]{@{}c@{}}Shad.\end{tabular} & $T_f$ & \begin{tabular}[c]{@{}c@{}}Indep.\end{tabular} & \begin{tabular}[c]{@{}c@{}}Shad.\end{tabular} \\ \Xhline{1pt}   
\multirow{9}{*}{RAD=5} & \multirow{3}{*}{C10} & 0.020 & 0.98 & 0.99 & 0.239 & 0.06 & 0.99 & 0.396 & 0.04 & 0.99 \\ \cline{3-11} 
 &  & \textbf{0.006} & \textbf{0.50} & \textbf{0.60} & \textbf{0.204} & \textbf{0.02} & \textbf{0.98} & \textbf{0.388} & \textbf{0.02} & \textbf{0.98} \\ \cline{3-11} 
 &  & 0.005 & 0.20 & 0.36 & 0.167 & 0.00 & 0.97 & 0.359 & 0.00 & 0.92 \\ \cline{2-11} 
 & \multirow{3}{*}{C100} & 0.027 & 0.16 & 0.98 & 0.048 & 0.08 & 0.99 & 0.494 & 0.52 & 1.00 \\ \cline{3-11} 
 &  & \textbf{0.023} & \textbf{0.07} & \textbf{0.93} & \textbf{0.040} & \textbf{0.03} & \textbf{0.97} & \textbf{0.403} & \textbf{0.00} & \textbf{1.00} \\ \cline{3-11} 
 &  & 0.010 & 0.00 & 0.85 & 0.033 & 0.00 & 0.92 & 0.035 & 0.00 & 0.65 \\ \cline{2-11} 
 & \multirow{3}{*}{TI} & 0.014 & 0.06 & 0.99 & 0.025 & 0.58 & 1.00 & 0.107 & 0.03 & 0.99 \\ \cline{3-11} 
 &  & \textbf{0.011} & \textbf{0.01} & \textbf{0.98} & \textbf{0.013} & \textbf{0.00} & \textbf{0.99} & \textbf{0.097} & \textbf{0.02} & \textbf{0.94} \\ \cline{3-11} 
 &  & 0.009 & 0.00 & 0.97 & 0.010 & 0.00 & 0.93 & 0.085 & 0.00 & 0.92 \\ \Xhline{1pt}
\multirow{9}{*}{RAD=15} & \multirow{3}{*}{C10} & 0.201 & 0.40 & 0.83 & 0.530 & 0.23 & 0.99 & 0.654 & 0.09 & 0.99 \\ \cline{3-11} 
 &  & 0.197 & 0.40 & 0.82 & \textbf{0.449} & \textbf{0.01} & \textbf{0.99} & \textbf{0.644} & \textbf{0.01} & \textbf{0.99} \\ \cline{3-11} 
 &  & \textbf{0.076} & \textbf{0.00} & \textbf{0.82} & 0.435 & 0.00 & 0.98 & 0.643 & 0.00 & 0.98 \\ \cline{2-11} 
 & \multirow{3}{*}{C100} & 0.144 & 0.07 & 1.00 & 0.189 & 0.03 & 0.99 & 0.622 & 0.26 & 1.00 \\ \cline{3-11} 
 &  & \textbf{0.130} & \textbf{0.01} & \textbf{0.98} & \textbf{0.176} & \textbf{0.02} & \textbf{0.98} & \textbf{0.566} & \textbf{0.00} & \textbf{1.00} \\ \cline{3-11} 
 &  & 0.095 & 0.00 & 0.86 & 0.147 & 0.00 & 0.94 & 0.560 & 0.00 & 0.90 \\ \cline{2-11} 
 & \multirow{3}{*}{TI} & 0.087 & 0.23 & 0.99 & 0.028 & 0.10 & 0.98 & 0.267 & 0.04 & 1.00 \\ \cline{3-11} 
 &  & \textbf{0.063} & \textbf{0.03} & \textbf{0.99} & \textbf{0.024} & \textbf{0.06} & \textbf{0.94} & \textbf{0.244} & \textbf{0.01} & \textbf{0.98} \\ \cline{3-11} 
 &  & 0.059 & 0.00 & 0.98 & 0.021 & 0.00 & 0.90 & 0.225 & 0.00 & 0.95 \\ \Xhline{1pt}
\end{tabular}}
\end{table}

\vspace{-1ex}

\section{Conclusion}

As a novel concept, fingerprinting DNNs is a promising way to verify the IP of the DNN model via its natural features without modifying the model. 
In this paper, instead of fingerprinting DNN models via its predictions, we give an effective DNN fingerprinting method via inference time to protect the IP for multi-exit models. 
Comprehensive evaluations show that our method outperforms the state-of-the-art DNN fingerprinting method considering the uniqueness and robustness.

\newpage
\bibliography{iclr2022_conference}
\bibliographystyle{iclr2022_conference}

\newpage
\appendix
\section{Experimental details}
\label{appendix:details}

\textbf{Training details. }
We trained models, especially the independent ones, on GPUs of type Nvidia 1080Ti, TITAN Xp, and RTX 3090 equipped with CUDA 11.1, and using the framework Pytorch 1.9.0.
We adopt the following training setting for all datasets: the backbone CNN model is trained with SGD optimizer with initial learning rate 0.01, momentum 0.9, weight decay $5\times10^{-4}$ for VGG-16 and $1\times10^{-1}$ otherwise.
The learning rate is decayed by factor $\gamma=0.1$ at epoch 20 and 40 for training on CIFAR-10 and at epoch 40 and 70 for other datasets.
To transform a CNN into an SDN, we add and optimize the ICs with Adam optimizer of learning rate 0.001 decayed by 0.1 at epoch 15.  
We also apply the random horizontal flip and the random crop to avoid overfitting.
To craft compressed models, we use Pytorch to quantize and prune SDNs.
As for normal finetuning, we use the same learning rate at the end of the training process, \ie 0.001, with Adam~\cite{kingma2014adam} optimizer.
We use PGD attack for adversarial training (AT), with 10 iterations and perturbation constraint $\epsilon=8/256$.
The hyperparameters (\eg, learning rate) remain the same as normal finetuning.


\textbf{Fingerprinting details. }
Our fingerprints are crafted with $c=10$ in Equation~\ref{eq:cw_optimize} and the loss is optimized by Adam optimizer of $0.01$ learning rate.
The optimization steps are 2000 for ResNet-56 and 1000 for the others, because ResNet-56 has double internal exits of the other architectures, rendering it more difficult to craft fingerprints.
The model inference times are measured on a single Nvidia RTX 3090 GPU.
As for the AE-based fingerprinting~\cite{conferrable}, the iteration step is 500 and the learning rate is 0.01 because it is sufficient to reduce the loss to 0.

\section{Experimental details on limitations of baseline}
\label{appendix:limitation_aebased}

In Figure~\ref{fig:adversarial_finetuning}, we first finetune a CNN of ResNet-56 trained on CIFAR-100 with PGD-based AT for 10 epochs. 
Then, we transform the CNN modified by At into an SDN.
Since AT can enhance the model's robustness against adversarial examples, the fingerprinting accuracy (FP accuracy) is close to zeros.
Therefore, unlike normal fingerprinting, the AE-based fingerprints are not able to detect suspicious models modified by AT.
On the other hand, the introduction of ICs can compensate for the loss of accuracy caused by AT, so we can observe the accuracy score (orange) of prediction under RAD=5 is close to the SDNs whose backbone CNNs are modified by normal finetuning.

\section{Additional experimental results}
\label{appendix:additional}

\begin{table}[!htbp]
\centering
\caption{Average EEC AUC scores of benign samples for the target multi-exit model. }
\label{tab:eec_score}
\resizebox{0.68\linewidth}{!}{
\begin{tabular}{c|c|c|c|c|c|c|c|c|c}
\Xhline{1pt}
\multirow{2}{*}{\textbf{Exit Threshold}} & \multicolumn{3}{c|}{\textbf{ResNet-56}} & \multicolumn{3}{c|}{\textbf{VGG-16}} & \multicolumn{3}{c}{\textbf{MobileNet}} \\ \cline{2-10} 
 & C10 & C100 & TI & C10 & C100 & TI & C10 & C100 & TI \\ \Xhline{1pt}
 RAD=5 & $0.03$ & $0.26$ & $0.25$ & $0.77$ & $0.72$ & $0.35$ & $0.84$ & $0.89$ & $0.65$ \\ \hline
 RAD=15 & $0.55$ & $0.52$ & $0.49$ & $0.91$ & $0.86$ & $0.60$ & $0.95$ & $0.95$ & $0.82$ \\ 
 \Xhline{1pt}
 \end{tabular}
 }
 \end{table}

\begin{table}[!htbp]
\vspace{-2em}
\centering
\caption{Average EEC AUC for 100 models of IC retraining + pruning (rate 0.2).}
\label{tab:icretrain_prune0.2_eec}
\resizebox{0.68\linewidth}{!}{
\begin{tabular}{c|c|c|c|c|c|c|c|c|c}
\Xhline{1pt}
\multirow{2}{*}{\textbf{RAD}} & \multicolumn{3}{c|}{\textbf{ResNet-56}} & \multicolumn{3}{c|}{\textbf{VGG-16}} & \multicolumn{3}{c}{\textbf{MobileNet}} \\ \cline{2-10} 
 & C10 & C100 & TI & C10 & C100 & TI & C10 & C100 & TI \\ \Xhline{1pt}
RAD=5 & 0.00 & 0.00 & 0.00 & 0.00 & 0.00 & 0.00 & 0.05 & 0.06 & 0.07 \\ \hline
RAD=15 & 0.00 & 0.01 & 0.00 & 0.00 & 0.00 & 0.00 & 0.26 & 0.21 & 0.10 \\ \Xhline{1pt}
\end{tabular}}
\end{table}

\begin{table}[!htbp]
\vspace{-2em}
\centering
\caption{Average EEC AUC for 100 models of IC retraining + pruning (rate 0.4).}
\label{tab:icretrain_prune0.4_eec}
\resizebox{0.68\linewidth}{!}{
\begin{tabular}{c|c|c|c|c|c|c|c|c|c}
\Xhline{1pt}
\multirow{2}{*}{\textbf{RAD}} & \multicolumn{3}{c|}{\textbf{ResNet-56}} & \multicolumn{3}{c|}{\textbf{VGG-16}} & \multicolumn{3}{c}{\textbf{MobileNet}} \\ \cline{2-10} 
 & C10 & C100 & TI & C10 & C100 & TI & C10 & C100 & TI \\ \Xhline{1pt}
RAD=5 & 0.00 & 0.00 & 0.00 & 0.00 & 0.00 & 0.00 & 0.08 & 0.11 & 0.09 \\ \hline
RAD=15 & 0.01 & 0.06 & 0.02 & 0.05 & 0.00 & 0.00 & 0.38 & 0.33 & 0.12 \\ \Xhline{1pt}
\end{tabular}}
\end{table}

\begin{table}[!htbp]
\vspace{-2em}
\centering
\caption{Average EEC AUC for 100 models of IC retraining + quantization.}
\label{tab:icretrain_quant_eec}
\resizebox{0.68\linewidth}{!}{
\begin{tabular}{c|c|c|c|c|c|c|c|c|c}
\Xhline{1pt}
\multirow{2}{*}{\textbf{RAD}} & \multicolumn{3}{c|}{\textbf{ResNet-56}} & \multicolumn{3}{c|}{\textbf{VGG-16}} & \multicolumn{3}{c}{\textbf{MobileNet}} \\ \cline{2-10} 
 & C10 & C100 & TI & C10 & C100 & TI & C10 & C100 & TI \\ \Xhline{1pt}
RAD=5 & 0.02 & 0.02 & 0.02 & 0.00 & 0.00 & 0.01 & 0.09 & 0.14 & 0.14 \\ \hline
RAD=15 & 0.02 & 0.03 & 0.02 & 0.00 & 0.01 & 0.02 & 0.33 & 0.32 & 0.17 \\ \Xhline{1pt}
\end{tabular}}
\end{table}

\clearpage

\begin{figure}[!htbp]
    \centering
    \includegraphics[width=\textwidth]{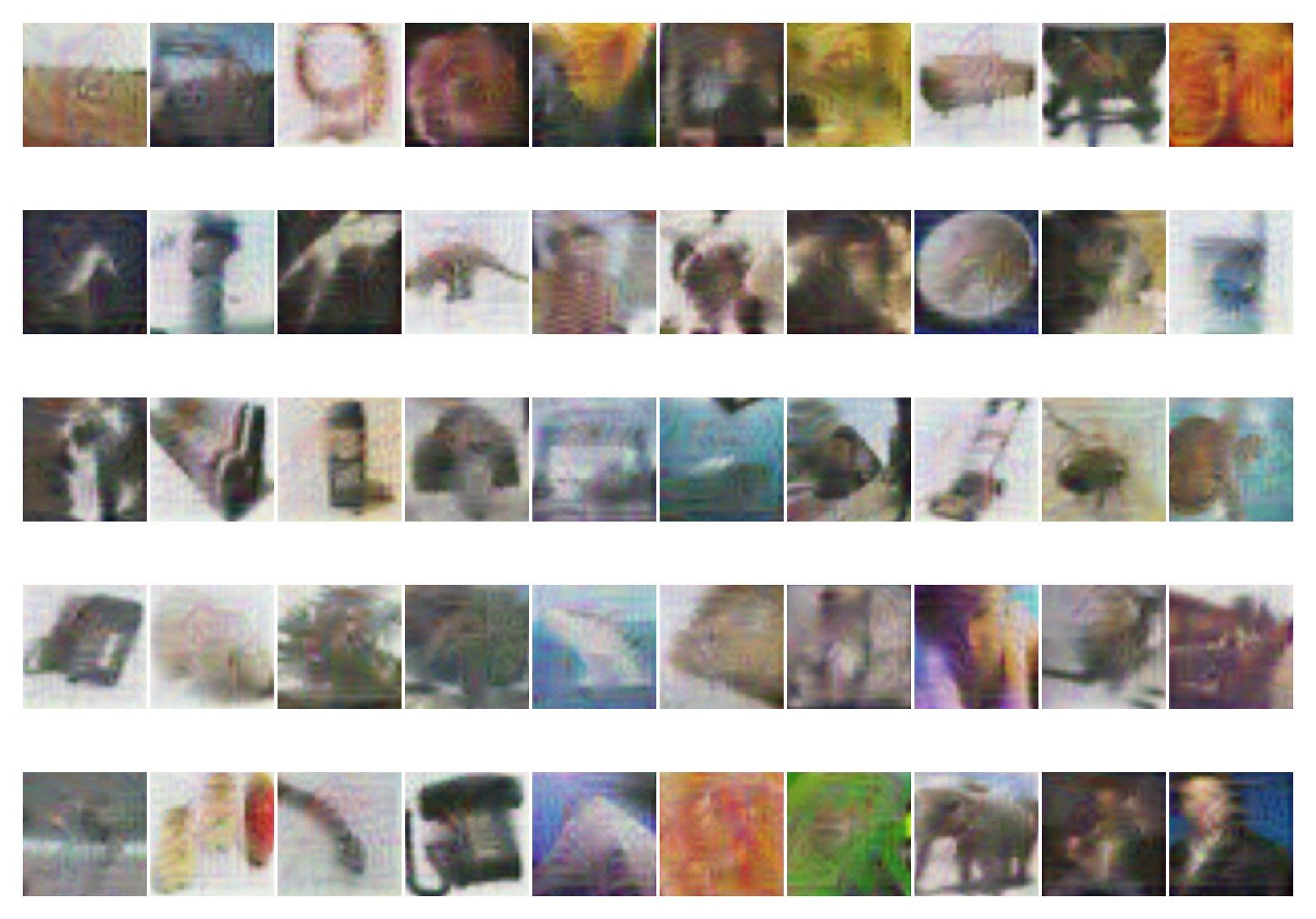}
    \vspace{-5ex}
    \caption{Fingerprint samples for multi-exit MobileNet model on CIFAR-100.}
    \label{fig:fingerprint_example_cifar100}
\end{figure}

\end{document}